\documentclass[aps,prb,preprint,showpacs,superscriptaddress,floatfix,nofootinbib]{revtex4-1}
\usepackage{graphicx,graphics}
\usepackage{dcolumn}
\usepackage{amsmath,amssymb,amsfonts}
\usepackage{latexsym,verbatim}
\usepackage{bm}
\usepackage{color}
\usepackage[breaklinks=true,colorlinks,citecolor=blue,linkcolor=blue,urlcolor=blue]{hyperref}
\usepackage{tabularx}

\newcommand{\be}{\begin{equation}}
\newcommand{\ee}{\end{equation}}
\newcommand{\bea}{\begin{eqnarray}}
\newcommand{\eea}{\end{eqnarray}}

 \newcommand{\bwt}{\begin{widetext}}
\newcommand{\ewt}{\end{widetext}}

\DeclareMathOperator{\sign}{sign}

\newcommand{\bi}{\begin{itemize}}
\newcommand{\ei}{\end{itemize}}
\newcommand \Disc {\mbox{Disc}}
\newcommand \tr {\mbox{{\bf Tr}}}

\begin{document}

\title{Unveiling Topological Modes on Curved Surfaces}

\author{Dmitry S. Ageev}
\email{ageev@mi-ras.ru}
\affiliation{Department of Mathematical Methods for Quantum Technologies, Steklov Mathematical Institute of Russian Academy of Sciences, Gubkin str. 8, 119991 Moscow, Russia}

\author{and Askar A. Iliasov}
\email{a.iliasov@science.ru.nl}
\affiliation{Institute for Molecules and Materials, Radboud University, Heyendaalseweg 135, 6525AJ Nijmegen, \mbox{The Netherlands}}

\begin{abstract}
In this paper, we investigate topological modes of different physical systems defined on arbitrary two-dimensional curved surfaces. We consider the shallow water equations, inhomogeneous Maxwell's equations,  Jackiw-Rebbi model and show how the topological protection mechanism responds to the presence of curvature in different situations. We show the existence of a line gap in the considered models and study the condition on the curve, which can host topological modes.
\end{abstract}

\maketitle

\section{Introduction}

Topological insulators have become an active area of research in recent years due to their fascinating properties and potential applications in various fields, such as electronics, photonics, and spintronics~\cite{Kane:2010,Moore:2010}. These materials are characterized by their unique topological properties, which give rise to protected edge states that are robust against disorder and impurities~\cite{Ludwig:2010}. In recent years, the underlying topological mechanisms have also been discovered in the non-standard physical setups, for example in the context of classical mechanics, hydrodynamics, electric circuits, plasma waves, and even game theory~\cite{Irvine:2015, Xuetal:2017, FuQin:2021, Parker:2020, Parker:2021, Vitelli:2022, Huber:2016, PanSun:2022, Frey:2020}.

An important example of such a system on the curved spacetime with topological protection mechanism is the well-known equatorial waves, such as Kelvin and Yanai waves. It was shown, that they are topologically trapped by the ``interface'' between the two hemispheres of Earth~\cite{Delplace:2017}. This observation has paved the way for new studies of topologically protected waves on surfaces in various physical systems~\cite{Efimkin:2022, Finnigan:2022, Green:2020} and gauge formulation of hydrodynamic equations \cite{Tong:2023, Nastase:2023, Eling:2023}. In general most research on topological materials has focused on flat spaces, while it is known that curvature (real or effective) may play an important role in the study of an electronic system~\cite{Guinea:2012, Dai_etal:2019,Fasolino:2007}.

In this paper, we explore the physics of topological matter and waves for various models, with a focus on curvature effects. These models include the shallow water equations, Maxwell's, and the Dirac equations on curved surfaces. We discuss the interplay between non-zero curvature and non-trivial band topology. In~\cite{Green:2020}, various cases of physical systems on surfaces of revolution were considered, and it was shown how topologically protected modes arise near geodesics. For a general surface, the curvature turns the initially Hermitian system to be non-Hermitian (namely for the shallow water and Maxwell's equations), which seems to be different from examples considered in~\cite{Green:2020} (where authors focused on the surfaces of revolution and neglect non-Hermitian terms). 

Topological states in non-Hermitian systems have attracted much attention recently, significantly extending the viable examples of topological matter ~\cite{nherm1,nherm2,nherm3,Ashida:2020, Shankar:2017, Okuma:2023, Kunst:2019, Bergholtz:2021}. We show the presence of a line gap in the bands corresponding to the models under consideration, which is the key property of the existence of topologically protected states. It turns out that for the Jackiw-Rebbi model, the change of sign of a chiral term is enough, while in other cases, there are some additional conditions for the chiral force.
 
This paper is organized as follows. In section~\ref{sec:setup}, we comment on the general type of physical systems we  consider. In section~\ref{sec:bands}, we study the shallow water,  Maxwell's equations in ingomogeneous medium, and the Jackiw-Rebbi model  for different setups  defined on curved surfaces in the context of topologically protected modes in these systems.

\section{General setup}\label{sec:setup}

Generally, we consider some system~${\cal A}$ defined on a two-dimensional spatial manifold $M$ with some coordinate system. Now let us describe the following objects which will be needed later:
\begin{itemize}
\item ``Physical observables'': the set of functions $f_i = f_i(x_1, x_2)$ defined on this manifold. These functions depend on the particular physical system we consider --- they can be different hydrodynamical parameters of the fluid, such as velocity, vorticity, height functions in shallow water, different parameters of plasma, shear in continuous media, diffusion constants varying in space and time, etc. 
\item The set of differential equations 
$F_j\left[\partial_t f_i, f_i, \nabla f_i, {\mathfrak{c} }(\Omega \land \mathfrak{f})\right]=0$  depending on the time derivatives of~$f_i$, covariant derivatives of~$f_i$ defined with respect to~$M$ and some function ${\mathfrak{c}}(\Omega \land \mathfrak{f})$ depending on the vector fields $\Omega$ (corresponding to some external source acting on our system) and~$\mathfrak{f}$. The components of the vector fields~$\mathfrak{f}$ are assumed to depend on the functions~$f_i$ and their derivatives. The field~${\mathfrak{c}}$ corresponds to some ``chiral force''. For example, in the Kelvin-Yanay topological waves, it is the Coriolis force created by the interplay of the velocity field $\mathfrak{f}=\mathfrak{v}$ and external force defined by the rotation of Earth. In general, we assume that the normal component of~$\Omega$ and, hence, the chiral force~${\mathfrak{c}}$ vanish along some curve $\gamma$ defined on~$M$.
\end{itemize}

In the examples we are going to consider (the shallow water equations, Maxwell's equations in inhomogeneous medium, the Jackiw-Rebbi model for curved interfaces), the curve~$\gamma$ will correspond to the location where the topological protection effects emerge (for example, topologically protected waves propagating along the equator). In~\cite{Green:2020}, it was shown how topological modes emerge on the surface of revolution naturally admitting~$U(1)$ symmetry, which includes the shallow water topologically protected modes of~\cite{Delplace:2017}. In this manuscript we consider how topologically protected modes arise on {\it arbitrary} general two-dimensional spatial manifolds, thus extending the previous results of~\cite{Green:2020}.

$\,$

Finally, let us notice that, an important ingredient of our consideration will be the special coordinate chart --- Fermi coordinates. Consider the coordinates around a curve~$\gamma$ as follows: the coordinate~$x$ corresponds to the direction along~$\gamma$, and the coordinate~$y$ parametrizes the transversal direction, i.e., let $(x,y)$ be  the Fermi coordinates. It is convenient to use them  to perform the linearization of the equation defining the system in the vicinity of curve ~$\gamma$  and proceed with calculating the Chern numbers for the corresponding bands via Fourier transform.

There is no general procedure for obtaining exact analytical expressions for the Fermi coordinates (see appendix~\ref{sec:fermi_cooordinates} for some explicit examples). However, it is well known that it is possible to write down a well-defined series expansion when~$\gamma$ is a geodesic~\cite{Manasse:1963, Dey:2014}.  For a two-dimensional manifold with the metric~$g_{ij}$, the expansion to the second order has the form\footnote{We use that $R_{yyyy}=R_{xyyy}=0$ and $R_{xyxy}=K/\det g_{ij}$ for 2D surfaces.}
\begin{gather}\label{eq:fermi_expansion}
    g_{xx}\approx1-K y^2, \\
    g_{xy}\approx 0 \\
    g_{yy} \approx 1
\end{gather}
and for Christoffel symbols, the corresponding expansion to the first order is
\begin{gather}\label{eq:fermi_expansion_Gamma}
    \Gamma^{x}_{xy}=\Gamma^{x}_{yx}\approx -K y\\
    \Gamma^{x}_{yy}\approx Ky
\end{gather}
where all other Christoffel symbols are zero, and $K$ is the Gaussian curvature. \footnote{ It is also interesting to note that the dependence on the coordinate $x$ along the geodesics is included only via the curvature.}.

\section{Bands and spectrum in curved spacetime:}\label{sec:bands}
\subsection*{Shallow water equations}

The first prototypical example we  consider is the shallow water equations,
\begin{gather}\label{eq:hydroeq}
    \partial_t h+\nabla\cdot(h{\bf u})=0, \\
    (\partial_t+{\bf u}\cdot\nabla){\bf u}=-g\nabla h+{\bf\Omega}\times {\bf u},
\end{gather}
which describes the dynamics of a fluid column.

 Here, ${\bf\Omega} \times {\bf u}$ is the chiral force acting on the fluid flow, and in the shallow water equations, it is just the local Coriolis force acting on the fluid. 

In order to apply topological arguments, we consider a linearized version of Eq. \eqref{eq:hydroeq}. We expand the shallow water equations around the rest state $(h,{\bf u})=(\langle h \rangle, {\bf 0})$. Having done that, we obtain the linearization of \eqref{eq:hydroeq} in the form
  \begin{gather}\label{eq:hydroeq_2}
    \partial_t h+\langle h\rangle\nabla {\bf u}=0, \\
    \partial_t {\bf u} =-g\nabla h+{\bf\Omega}\times {\bf u},
\end{gather}
  The next step is to perform Fourier transform and re-write Eq. \eqref{eq:hydroeq_2} as an effective stationary Shr{\"o}dinger equation:
 \begin{gather}\label{eq:shallow_water}
    i\omega\begin{pmatrix} h\\ u_x\\u_y\end{pmatrix}=\begin{pmatrix} 0 & \langle h\rangle(-i k_x +\Gamma_x)& \langle h\rangle(-i k_y+\Gamma_y) \\ -i g k_x &0& -\Omega_n\\-i g k_y & \Omega_n & 0\end{pmatrix}\begin{pmatrix} h\\ u_x\\u_y\end{pmatrix},
\end{gather}

where~$\Omega_n$ is a component of the vector~${\bf \Omega}$ that is normal to the surface. By~$\Gamma_x$ and~$\Gamma_y$ we denote the following combination of the two Christoffel symbols
\begin{gather} 
    \Gamma_x=\Gamma^{x}_{xx}+\Gamma^{y}_{xy}, \\ \Gamma_y=\Gamma^{x}_{yx}+\Gamma^{y}_{yy},
\end{gather}
where the Einstein summation convention is not assumed.

$\,$

For the shallow water equations given by~\eqref{eq:shallow_water}, the vector~$\Omega_n$ and the chiral force defined by~$\Omega_n$ are equal to zero on the line~$\gamma$, by definition. It is a parameter of the linearized equation along with~$\Gamma_x$ and~$\Gamma_y$. Equations~\eqref{eq:shallow_water} correspond to the equations for the topological equatorial waves with an additional curvature term (here we follow the calculations from~\cite{Delplace:2017}). The dispersion relation defined by~\eqref{eq:shallow_water} for the ``equatorial waves bands'' is given by
\be\label{eq:disp-shallow-1}
    \omega^3-\omega\left[\Omega^2+c^2 (k^2+ik_x\Gamma_x+ik_y\Gamma_y\right)]+c^2 \Omega_n\left[k_y\Gamma_x-k_x\Gamma_y\right]=0.
\ee
where $c=\sqrt{g\langle h\rangle}$ is the speed of shallow water gravity waves. The solutions of the dispersion equation are complex-valued and, therefore, explicitly demonstrate the non-Hermiticity.

$\,$

Despite the non-Hermiticity, one can argue that topologically protected modes are still present in the model described by~\eqref{eq:shallow_water}. Generally speaking, the key point for the consideration of non-Hermitian systems like~\eqref{eq:shallow_water} is the presence of a line gap --- the possibility of a clear separation of the spectrum in the bands on ``disconnected'' components. In other words, there is a curve separating two bands in the spectrum (in the complex plane)~\cite{Delplace:2022}. Given the presence of the line gap one can argue that the existence of the Chern number generalization which formally coincides with the ordinary Chern numbers (i.e. when the system is Hermitian. In our case, it is when the curvature is absent). In what follows, we use the term ``Chern number'' and ``generalized Chern number'' interchangeably.

$\,$

One can show that for a non-Hermitian system, there is a formula relating the Chern numbers~$C$ and the projection operators~$P$:
\begin{equation}\label{eq:NH_Chern}
    C=\frac{1}{2\pi i}\int \tr(P dP \wedge dP),
\end{equation}
where~$P$ is given by $P=|\psi\rangle \langle \tilde \psi|$, $|\psi\rangle$ is the right eigenvector of the Hamiltonian and $\langle \tilde \psi|$ is the left eigenvector of the Hamiltonian. The projector is non-orthogonal, and for Hermitian systems, it reduces to the ordinary orthogonal projector $P=|\psi\rangle \langle \psi|$. The (generalized) Chern number given by this formula is related by a homotopy, with respect to some parameter, to the Chern number of the Hermitian system~\cite{Delplace:2022}. Also, it is important to notice that the shallow water equations, in fact, do not possess well-defined topological numbers in the flat case \cite{Rossi:2023}. As was shown in~\cite{Tauber:2019}, to calculate the Chern numbers correctly, one should also properly regularize the chiral term, for example, by addition of the odd-viscosity term $\epsilon k^2$, where~$\epsilon$ is the viscosity amplitude. 

$\,$

We show in Appendix~\ref{sec:line-gap} that the model given by~\eqref{eq:shallow_water} contains the line gap if (here we restore the rescaling by $c$):
\be \label{eq:line_gap_cond_1}
\sqrt{\Gamma^2_x+\Gamma^2_y}<\frac{|\Omega_n|}{3c}.
\ee 
 Notice that there are two important implications of this inequality. At first, we can conclude that $\Gamma_x$ and $\Gamma_y$ must vanish on the curve $\gamma$, where $\Omega_n$ equals zero. From the expansion of the metric in the vicinity of a geodesic \eqref{eq:fermi_expansion}, it follows that Christoffel symbols are zero on a geodesic. In other words, $\gamma$ should be a geodesic to support the line gap.

The second condition for the existence of line gap is the requirement on the derivative of $\Omega_n$ at $\gamma$. Since Christoffel symbols are linear functions of coordinates in the vicinity of a geodesic, the derivatives of $\Omega_n$ should be larger than the derivatives of Christoffel symbols. Using the linear approximation for the chiral term $\Omega_n=\partial_y \Omega_n y$ and the expansion for the Fermi coordinates $\Gamma^x_{xy}=-K y$, we obtain
\be\label{eq:line_gap_cond_2}
|K|<\frac{|\partial_y \Omega_n|}{3c}
\ee
where the derivative of the chiral term is calculated in Fermi coordinates, and $K$ is the Gaussian curvature.

As an example, let us consider shallow water equations on a sphere rotating with angular velocity ${\bf\Omega_0}$. The chiral term $\Omega_n$ changes sign on the equator, which is a geodesic, and the Fermi coordinates are given by the rescaled spherical coordinates. The Gaussian curvature and the derivative of chiral term are $K=R^{-2}$ and $|\partial_y \Omega_n|=2|{\bf\Omega_0}|/R$, where $R$ is the radius of the sphere. Using the condition \eqref{eq:line_gap_cond_2}, we can conclude that there is a line gap only if the rotation is fast enough 

\be |{\bf\Omega_0}|>\frac{3}{2} \frac{\sqrt{g\langle h\rangle}}{R}
\ee 
If the line gap is present, the existence of the topological modes and their analysis of the Chern numbers follow from the bulk-boundary correspondence \cite{Murakami:2016, Kawabata:2019, Yao:2018, Schindler:2023}.

\subsection*{Maxwell's equations in inhomogeneous medium}

Another example is macroscopic Maxwell's equations on a two-dimensional surface with inhomogeneous dielectric coefficients~\cite{Delplace:2020}.
Maxwell's equations in two-dimensional non-dispersive media take the form~\cite{Mechelen:2018}
\begin{gather}\label{eq:maxw}
    \begin{cases}
        \partial_t D_i-\epsilon_{ij}\partial_{j} H_z=0, \\
        \partial_t B_z-\partial_y E_x+\partial_x E_y -\Gamma^{x}_{xy}E_x+\Gamma^{x}_{xx}E_y=0,
    \end{cases} 
\end{gather}
where curvature effects are incorporated in the action of the covariant derivative on the dual electromagnetic tensor (see Appendix~\ref{sec:maxwell}).
 
Here, $\epsilon_{ij}$ is the Levi-Civita symbol, and vectors $(E_x, E_y, B_z)$ and $(D_x, D_y, H_z)$ are related by the linear response function~$\mathcal{M}$ 
\begin{gather}
   \begin{pmatrix}
        D_x\\ D_y\\B_z
   \end{pmatrix}= \begin{pmatrix}
\varepsilon_{xx}&\varepsilon_{xy}&\chi_x\\\varepsilon_{yx}&\varepsilon_{yy}&\chi_y\\\chi^{*}_x&\chi^{*}_y&\mu
   \end{pmatrix}  \begin{pmatrix}
        E_x\\ E_y\\H_z
   \end{pmatrix}.
\end{gather}
After the Fourier transform, equations~\eqref{eq:maxw} can be re-written again as a ``generalized'' eigenvalue problem for the operator~$H_0(k)$, 
\begin{equation} \label{eq:H0}
    H_0(k)f=\omega \mathcal{M}(k,\omega)f,
\end{equation}
where $f=(E_x,E_y,H_z)^{T}$. After some algebra, one can write down it explicitly as
\begin{gather}
\begin{pmatrix}0&0&-k_y\\0&0&k_x\\-k_y-i\Gamma^{x}_{xy}&k_x+i\Gamma^{x}_{xx}&0
   \end{pmatrix}  \begin{pmatrix}
        E_x\\ E_y\\H_z
   \end{pmatrix}= \omega \begin{pmatrix}
\varepsilon_{xx}&\varepsilon_{xy}&\chi_x\\\varepsilon_{yx}&\varepsilon_{yy}&\chi_y\\\chi^{*}_x&\chi^{*}_y&\mu
   \end{pmatrix}  \begin{pmatrix}
        E_x\\ E_y\\H_z
   \end{pmatrix}.
\end{gather}

A particularly interesting example is a  strongly magnetized plasma, i.e. when cyclotron frequency~$\omega_c$ is larger than the plasma frequency, $\omega_c \gg \omega_p$ and $\omega_c \gg\omega$. In this case, the equation~\eqref{eq:H0} reduces to
\begin{gather}
\begin{pmatrix}0&0&-k_y\\0&0&k_x\\-k_y-i\Gamma^{x}_{xy}&k_x+i\Gamma^{x}_{xx}&0
    \end{pmatrix}  \begin{pmatrix}
        E_x\\ E_y\\H_z
    \end{pmatrix}= \begin{pmatrix}
\omega &i\frac{\omega_c}{\omega_p}&0\\-i\frac{\omega_c}{\omega_p}&\omega &0\\0&0&\omega
    \end{pmatrix}  \begin{pmatrix}
        E_x\\ E_y\\H_z
    \end{pmatrix},
\end{gather}
and we can explicitly write down the dispersion relation
\be\label{eq:disp-maxwell-1}
    \omega^3-\omega\left[\left(\frac{\omega_c}{\omega_p}\right)^2+k^2+ik_x\Gamma^x_{xx}+ik_y\Gamma^{x}_{xy}\right]+\frac{\omega_c}{\omega_p}\left[k_y\Gamma^{x}_{xx}-k_x\Gamma^{x}_{xy}\right]=0.
\ee
We see that Maxwell's equations on a surface with a non-trivial inhomogeneous linear response function have the same form as the shallow water equations. Therefore, all the above considerations hold in the general situation except one peculiar feature. Instead of the combination $\Gamma_x$ and $\Gamma_y$ the dispersion relation now contains the Christoffel symbols themselves. A simple illustrative example where this could effect topological modes is a sphere.  The  Christoffel symbols present in \eqref{eq:disp-maxwell-1} equal zero, so the line gap in dispersion relation is present for all values of $\omega_c$.  This is not the case as we have seen for the shallow water equations, for which topologically protected modes propagate along geodesics.

\subsection*{Dirac equation in curved spacetime}

Now let us consider our last example --- the Dirac equation on a two-dimensional surface with an inhomogeneous mass term, i.e., the generalization of the Jackiw-Rebbi model in a curved spacetime. The chiral term is a bit different here compared to other cases since the mass term does not have a natural interpretation as a projection of a vector. As in the Jackiw-Rebbi model in the flat space, we assume that the mass term changes its sign on some line on the surface. In the momentum space the  Dirac equation on the curved spacetime has the form
\begin{gather}\label{eq:curved_Dirac}
   \begin{pmatrix}-\omega-m & -ik'_x-k'_y+\frac{i}{2}R_x+\frac{1}{2}R_y\\ -ik'_x+k'_y-\frac{i}{2}R_x+\frac{1}{2}R_y & \omega-m \end{pmatrix}\begin{pmatrix}\psi_{x} \\ \psi_{y}\end{pmatrix}=0,
\end{gather}
where we assume that since the mass~$m$ has a sharp change across some line (and is constant locally). Introducing the rescaled wave-vectors $k'_{\mu}$ and setting up he notation
\begin{gather}
    k'_x=k_x e^{x}_{\bar x}+k_y e^{y}_{\bar x}, \quad
    k'_y=k_x e^{x}_{\bar y}+k_y e^{y}_{\bar y}, \\
    R_x=\omega^{\bar x\bar y}_x e^{x}_{\bar x}+\omega^{\bar x\bar y}_y e^{y}_{\bar x}, \quad R_y=\omega^{\bar x\bar y}_x e^{x}_{\bar y}+\omega^{\bar x\bar y}_y e^{y}_{\bar y}.
\end{gather} 
we find that the dispersion relation corresponding to the equation~\eqref{eq:curved_Dirac} has the form
\begin{equation}
    \omega=\pm\sqrt{m^2+\left(k'_x+\frac{i R_y}{2}\right)^2+\left(k'_y-\frac{i R_x}{2}\right)^2}.
\end{equation}
It is worth noticing that the dispersion relation is similar to that of the shallow water equations. The wave vectors are effectively shifted by imaginary values of spin connection terms, and one can see that the bands have an intersection only when $m=k'_x=k'_y=0$ (see appendix ~\ref{sec:line-gap}). Therefore, repeating the same logical steps, one can prove the existence of the line gap, thus correctly defining the Chern numbers. Here, as in the previous examples, one also needs to regularize the chiral term in order to correctly calculate topological bulk invariants. Usually, it is done by the additional term $k\epsilon^2$~\cite{Bal:2019}, with~$\epsilon$ being a regularization parameter. There is one topologically protected edge mode propagating along the line, where the mass changes the sign. In contrast with the shallow water equations, the curvature does not affect the existence of the line gap and, therefore, the presence of a topologically protected mode, even if the mass term changes its sign not in the vicinity of a geodesic.

\subsection*{Circular domain wall and Dirac equation}

In the preceding sections, we studied the general case with a curved manifold, where the line of zeros of the chiral force was required to coincide with a geodesic. However, as we have shown, some physical systems admit topologically protected modes without this requirement. The main ingredient is the ``line gapped modes'', and it is possible to obtain them when the line of the ``chiral force'' zeros is not a geodesic. For example, in the case of the Dirac equation (with non-zero mass), two bands have an intersection only when the chiral term equals zero. Therefore, we can expect the emergence of topological modes for any smooth enough curves, for which we can build Fermi coordinates.

In appendix~\ref{appendix:polarDirac}, we consider the Jackiw-Rebbi model with the circular domain wall, i.e., with the spatially dependent mass of the form
\be 
    m=m_0\frac{r-r_0}{|r-r_0|},
\ee 
and show that for a large enough radius~$r_0$, one can observe a topologically protected mode propagating along the circular domain boundary 
\begin{equation}
    \omega=\sign(m')\frac{k}{r_0}.
\end{equation}
We see that the bulk-boundary correspondence works well here, although the Dirac Hamiltonian is non-Hermitian. This result can be shown for more general curves using the quasi-classical approximation~\cite{HuXieZhu:2022, Drouot:2022}.

\section{Discussion}

In this paper, we have investigated the interplay between the topological edge modes and curved geometry for different physical setups, namely the shallow water equations, inhomogeneous Maxwell's equations, and the Dirac equation defined on curved manifold. We have considered the cases where edge states are formed by a chiral term breaking the symmetry of the system Hamiltonians. In the cases when the Hamiltonians are Hermitian, one can observe the classical bulk-boundary correspondence by calculating the Chern number in the regions with different sings of the chiral term and usually one can conclude the existence of topologically protected edge modes. The presence of curvature makes the system non-Hermitian, and therefore one has to argue the existence of the line gap in the band spectrum. 

In all considered cases, the Christoffel symbols appear as the terms breaking the Hermiticity of Hamiltonians, moving the band structure to the complex plane. If we consider a general case, there are intersections between modes, and it is a complicated problem to define topological invariants and establish the existence of the corresponding edge states. However, if one considers the case where the chiral term changes sign on a geodesic, the Christoffel symbols become small (effectively, in the vicinity of the geodesic, the surface is flat), and one can write an explicit condition for the existence of the line gap between modes.

We have seen that the curvature's effect on the bands' structure depends on underlying physical systems. In some systems, such as the shallow water equations, the necessary condition for the existence of the line gap is the concurrence of a geodesic and the line of sign change of the chiral term. For other systems, the mentioned condition is important only for some manifolds (Maxwell's equations) or is not necessary, as in the case of the Dirac equations, where the line gap is present independently on curvature.

\section*{Acknowledgements}

We would like to thank A.A. Bagrov, M.I. Katsnelson, V.P. Pushkarev, and A.I. Belokon for discussions and comments on this manuscript.

\appendix

\section{Existence of line gap in shallow water equations}\label{sec:line-gap}

In this appendix, we show the existence of a line gap in the bands of the shallow water equations. The starting point of our analysis is the equation~\eqref{eq:disp-shallow-1} defining the bands in our case
\be\label{eq:disp-shallow-app}
    \omega^3-\omega\left[\Omega^2+k^2+ik_x\Gamma_x+ik_y\Gamma_y\right]+\Omega_n\left[k_y\Gamma_x-k_x\Gamma_y\right]=0,
\ee
where we make rescaling of the wave vectors and Christoffel symbols by $c=\sqrt{g\langle h \rangle}$ in comparison with Eq. \eqref{eq:disp-shallow-1}.

The condition that some polynomial $P$ has multiple roots can be expressed as the condition on the discriminant of the polynomial
\begin{equation}
    \Disc_x(P)=0.
\end{equation}
In the case of a depressed cubic polynomial $x^3+px+q$, the discriminant is given by the formula
\begin{equation}\label{eq:Disc}
    \Disc_x(P)=-4p^3-27q^2.
\end{equation}
If we neglect the constant term in Eq. \eqref{eq:disp-shallow-app}, the condition~\eqref{eq:Disc} simplifies to the equation
\begin{equation}
    \Omega^2+k_x^2+k_y^2+ik_x\Gamma_x+ik_y\Gamma_y=0.
\end{equation}
Since all the parameters are real numbers, this equality holds only at the point $\Omega=k_x=k_y=0$, and it does not depend on the curvature terms. Therefore, we can conclude that the bands have a line gap.

In the case if we are far away from a geodesic, the condition~\eqref{eq:Disc} for the points of the bands' intersection has the form
\begin{gather}\label{eq:Disc_shallow_water}
\begin{cases}
    (\Omega^2+k^2)^3-3(\Omega^2+k^2)(k_x\Gamma_x+k_y\Gamma_y)^2+\frac{27}{4}\Omega^2 (k_y\Gamma_x-k_x\Gamma_y)^2=0, \\
    (k_x\Gamma_x+k_y\Gamma_y)^3-3(k_x\Gamma_x+k_y\Gamma_y)(\Omega^2+k^2)^2=0.
\end{cases}
\end{gather}
We are interested in the solutions for small $\Omega$ and aim to obtain them when our considerations lead to well-defined topological indices.

To begin with, we introduce new coordinates in momentum space: \footnote{Possible degeneracy of this coordinate transformation does not change the calculations' result due to continuous dependence on the parameters $(\Gamma_x, \Gamma_y, \Omega)$.}
\begin{gather}
u=\Gamma_x k_x+\Gamma_y k_y\\\nonumber
v=-\Gamma_y k_x+\Gamma_x k_y
\end{gather}
The Eq. \eqref{eq:Disc_shallow_water} take the form:
\begin{gather}
\begin{cases}
    (\Omega^2+\frac{1}{\gamma^2}(u^2+v^2))^3-3(\Omega^2+\frac{1}{\gamma^2}(u^2+v^2))u^2+\frac{27}{4}\Omega^2 v^2=0, \\
    u^3-3u(\Omega^2+\frac{1}{\gamma^2}(u^2+v^2))^3=0.
\end{cases}
\end{gather}
where $\gamma^2=\Gamma^2_x+\Gamma^2_y$.

From the second equation, we obtain that one of the two equations should be satisfied
\begin{gather}\label{eq:cond-app-1}
    u=0, \\
    \frac{1}{\sqrt{3}}u=\Omega^2+\frac{1}{\gamma^2}(u^2+v^2)
\end{gather}

Let us consider the case $u=0$. Then one can see that
\begin{equation}\label{eq:u=0}
    (\Omega^2+\frac{v^2}{\gamma^2})^3=\frac{27}{4}\Omega^2 v^2.
\end{equation}

We are interested in a condition with an intersection between the bands. The curvature $\gamma$ and the chiral $\Omega$ terms are parameters, and we a looking for a critical case when the functions on the left side and the right side of eq.~\eqref{eq:u=0} are equal at some point $v=v_0$ as well as their first derivative to make the junction smooth enough
\begin{gather}
    \begin{cases}
    (\Omega^2+\frac{v^2_0}{\gamma^2})^2=\frac{9}{4}\Omega^2\gamma^2, \\
    (\Omega^2+\frac{v^2_0}{\gamma^2})^3=\frac{27}{4}\Omega^2 v_0^2.
    \end{cases}
\end{gather}
After some algebra, we obtain the critical value of the curvature term
\begin{equation}\label{eq:crit_bound}
    |\gamma|=\frac{|\Omega|}{3},
\end{equation}
and if $|\gamma|<|\Omega|/3$ there is no solution of eq.~\eqref{eq:u=0}.

Let us consider the second case, $\frac{1}{\sqrt{3}}u=\Omega^2+\frac{1}{\gamma^2}(u^2+v^2)$, from~\eqref{eq:cond-app-1}. In this case, we obtain
\begin{gather}
    \begin{cases}
    u^3=\frac{27\sqrt{3}}{32}\Omega^2v^2, \\
    (u-\frac{\gamma^2}{2\sqrt{3}})^2+v^2=\gamma^2(\frac{\gamma^2}{12}-\Omega^2),
    \end{cases}
\end{gather}

One can notice that if $\gamma^2\le 12\Omega^2$, there are no real solutions to these equations, except the point $\Omega^2=\gamma^2=u=v=0$. Since we have already obtained the lower bound for the absence of bands' intersections \eqref{eq:crit_bound}, this lower bound holds for the initial equations \eqref{eq:Disc_shallow_water}. Therefore, we can conclude that the bands are line-gapped if $\sqrt{\Gamma^2_x+\Gamma^2_y}<|\Omega|/3$.

\section{Maxwell's equations on curved surface}\label{sec:maxwell}

In this appendix, we briefly remind the derivation of equations~\eqref{eq:maxw}. We start as usual with the vector potential $A$ and the electromagnetic tensor $F$
\begin{equation}
    A^{\mu}=(\phi, A_x, A_y),
\end{equation}
\begin{equation}
    F^{\mu\nu}=\partial^{\mu}A^{\nu}-\partial^{\nu}A^{\mu}.
\end{equation}
The electric and magnetic fields in a vacuum are defined as
\begin{equation}
    E_{i}=-\partial_i \phi-\partial_t A_i,
\end{equation}
\begin{equation}
    B_z=\partial_x A_y-\partial_y A_x.
\end{equation}
Maxwell's equations take the form
\begin{gather}
    \nabla_\mu F^{\mu\nu}=J^{\nu}, \\
    \nabla_\mu \tilde F^{\mu}=0,
\end{gather}
where $J^{\nu}$ is the 3-current density, $J^{\nu}=(\rho,J_x,J_y)$, and $\tilde F^{\mu}$ is the dual tensor (it is a vector in 2+1D) defined as
\begin{equation}
    \tilde F^{\mu}=\frac{1}{2}\sqrt{g}\epsilon^{\mu \nu \rho} F_{\nu\rho},
\end{equation}
where $\epsilon^{\mu \nu \rho}$ is the Levi-Civita symbol. The components of the dual tensor are given by
\begin{equation}
    \tilde F^{\mu}=\sqrt{g}(B_z,-E_x,E_y),
\end{equation}
with the current density satisfying the continuity equation
\begin{equation}
   \partial_t \rho+\nabla_i J^{i}=0.
\end{equation}
Now, we can write Maxwell's equations in terms of electric and magnetic fields
\begin{gather}
    \nabla_i E^{i}=\rho, \\
    \partial_y B_z-\partial_t E_x=J_x, \\
    -\partial_x B_z-\partial_t E_x=J_y, \\
   \partial_t B_z-\partial_y E_x+\partial_x E_y -\Gamma^{x}_{xy}E_x+\Gamma^{x}_{xx}E_y=0.
\end{gather}
In the last equation, we see the presence of curvature coming from the covariant derivative of the dual tensor. Also, notice that all of the Christoffel symbols, including that with time indices, vanish.

Since we describe the media, we need to take into account the polarization $P_i$ and magnetization $M_z$ densities defined by the expressions (using the continuity equation)
\begin{gather}
    \rho=-\nabla_i P^{i}, \\
    J_x=\partial_t P_i+\partial_y M_z, \\
    J_y=\partial_t P_i-\partial_x M_z.
\end{gather}
Then, we define the electric and magnetic displacement fields as
\begin{gather}
    D_i=E_i+P_i, \\
    B_z=H_z+M_z.
\end{gather}
We obtain Eq.~\eqref{eq:maxw} by substituting these expressions back into the equations.

\section{Dirac equation on curved surface}

At first, we write the Dirac equation on a curved two-dimensional surface~\cite{Gallerati:2019}
\be
    [i\bar{\gamma}^{\mu}\mathit{D}_{\mu}-m]\psi=0,
\ee
where $\bar{\gamma}^{\mu}$ are gamma-matrices (in a curved spacetime)
\be
    \bar{\gamma}^{\mu}=e^{\mu}_{a}\gamma^a,
\ee
and $\mathit{D}_{\mu}$ is the covariant derivative
\be
    \mathit{D}_{\mu}=\partial_{\mu}-\frac{i}{4}\omega^{ab}_{\mu}M_{ab},
\ee
where $\omega^{ab}_{\mu}$ is the spin connection and $M_{ab}$ are the generators of $SO(1,2)$ group
\be
    M_{ab}=\frac{i}{2}[\gamma^a,\gamma^b].
\ee
We choose $\gamma^a$ as follows
\be
    \gamma^{a}=\{\sigma_3,\, -i\sigma_1,\, -i\sigma_2\},
\ee
where $\sigma_i$ are the Pauli matrices, so that $\gamma^a$ satisfy the condition for Clifford algebra, $\{\gamma^a,\gamma^b \}=2\eta^{ab}$. The Dirac equation has the form
\begin{gather}
    \Big(-\omega\sigma_3 -i(k_x e^{x}_{\bar x}+k_y e^{y}_{\bar x})\sigma_1-i(k_x e^{x}_{\bar y}+k_y e^{y}_{\bar y})\sigma_2 + \\
    + \frac{1}{2}(\omega^{\bar x\bar y}_x e^{x}_{\bar x}+ \omega^{\bar x\bar y}_y e^{y}_{\bar x})\sigma_2-\frac{1}{2}(\omega^{\bar x\bar y}_x e^{x}_{\bar y}+\omega^{\bar x\bar y}_y e^{y}_{\bar y})\sigma_1 -m\Big)\psi=0,
\end{gather}
or being rewritten explicitly in matrix form,
\begin{gather}
    \begin{pmatrix}-\omega-m & -ik'_x-k'_y+\frac{i}{2}R_x+\frac{1}{2}R_y\\ -ik'_x+k'_y-\frac{i}{2}R_x+\frac{1}{2}R_y & \omega-m \end{pmatrix}\begin{pmatrix}\psi_{x} \\ \psi_{y}\end{pmatrix}=0,
\end{gather}
where we use the following shorthand notation
\begin{gather}
    k'_x=k_x e^{x}_{\bar x}+k_y e^{y}_{\bar x}, \quad k'_y=k_x e^{x}_{\bar y}+k_y e^{y}_{\bar y}, \\
    R_x=\omega^{\bar x\bar y}_x e^{x}_{\bar x}+\omega^{\bar x\bar y}_y e^{y}_{\bar x} \quad R_y=\omega^{\bar x\bar y}_x e^{x}_{\bar y}+\omega^{\bar x\bar y}_y e^{y}_{\bar y}.
\end{gather}

\section{Dirac equation on plane with sharp mass change}\label{appendix:polarDirac}

We start with the Dirac equation described in the previous section of the appendix. In the case of polar coordinates, the vierbein is diagonal:
\be
    e^{\bar t}_{t}=1, \quad e^{\bar r}_{r} =1, \quad e^{\bar\phi}_{\phi}=r,
\ee
and the only non-zero coefficients of the spin connection are
\be
    \omega^{\bar\phi\bar r}_{\phi}=-\omega^{\bar r \bar\phi}_{\phi}=1.
\ee
Then, using the equation~\eqref{eq:curved_Dirac}, we can write down the Dirac equation
\begin{gather}
    \begin{pmatrix}-\omega-m & \partial_{r}-\frac{k}{r}+\frac{1}{2r}\\ \partial_{r}+\frac{k}{r}+\frac{1}{2r} & \omega-m \end{pmatrix}\begin{pmatrix}\psi_{r} \\ \psi_{\phi}\end{pmatrix}=0,
\end{gather}
where we shortly denote the wave vector along the angular coordinate as $k$, and the mass term $m(r)$ depends only on the radial coordinate. The curvature terms can be eliminated by the redefinition
\begin{gather}
    \begin{pmatrix}\psi_{r} \\ \psi_{\phi}\end{pmatrix}= r^{-\frac{1}{2}}\begin{pmatrix}u_{r} \\ u_{\phi}\end{pmatrix},
\end{gather}
and we have the equations
\begin{gather}
    \begin{pmatrix}-\omega-m & \partial_{r}-\frac{k}{r}\\ \partial_{r}+\frac{k}{r} & \omega-m \end{pmatrix}\begin{pmatrix}u{_r} \\ u_{\phi}\end{pmatrix}=0.
\end{gather}

To obtain bound modes, we need to solve these equations with the conditions $\psi(\infty)=0$ and regularity at $0$, as well as the normalization condition.

We obtain the following differential equations for $u_r$ and $u_{\phi}$
\begin{equation}
    \left[\partial^2_r -\frac{k(k+1)}{r^2}+(\omega^2-m^2)\right]u_r=0,
\end{equation}
\begin{equation}
    \left[\partial^2_r -\frac{k(k-1)}{r^2}+(\omega^2-m^2)\right]u_\phi=0,
\end{equation}
As an example, we consider the case in which the mass term sharply changes its sign at some radius $r_0$
\begin{gather}
    m = 
    \begin{cases}
        m_0,\, r<r_0, \\
        -m_0,\, r>r_0.
    \end{cases}
\end{gather}

In general, there are three cases, $\omega^2=m^2_0$, $\omega^2>m^2_0$, and $\omega^2<m^2_0$. For the simplicity of the analysis, without loss of generality, let us restrict ourselves to $k>\frac{1}{2}$.

Let us consider the case $\omega^2=m^2$. The general solution is given by
\begin{gather}
    \psi_r = 
    \begin{cases}
        ar^{\frac{1}{2}+k}+br^{-\frac{1}{2}-k},\, r<r_0, \\
        cr^{\frac{1}{2}+k}+dr^{-\frac{1}{2}-k},\, r>r_0.
    \end{cases}.
\end{gather}
Since we need a bounded solution, we should set $b=c=0$, therefore
\begin{gather}
    \psi_r = 
    \begin{cases}
        ar^{\frac{1}{2}+k},\, r<r_0, \\
        dr^{-\frac{1}{2}-k},\, r>r_0.
    \end{cases}
\end{gather}
However, if we substitute this into the original matrix equation, we obtain the conditions
\begin{equation}
    a(2k+1)r^{k-\frac{1}{2}}=0,
\end{equation}
which can be satisfied only if $a=0$, and from the continuity condition, we deduce that $d=0$; thus, we conclude that there are no bounded solutions for $\omega^2=m^2_0$.

In the case $\omega^2>m^2_0$, we have
\begin{gather}
    \psi_r = 
    \begin{cases}
        aJ_{k+\frac{1}{2}}\left(r\sqrt{\omega^2-m^2_0}\right)+bY_{k+\frac{1}{2}}\left(r\sqrt{\omega^2-m^2_0}\right),\, r<r_0, \\
        cJ_{k+\frac{1}{2}}\left(r\sqrt{\omega^2-m^2_0}\right)+dY_{k+\frac{1}{2}}\left(r\sqrt{\omega^2-m^2_0}\right),\, r>r_0,
    \end{cases}
\end{gather}
where $J_a(x)$ and $Y_a(x)$ are the Bessel functions of the first and second kinds, respectively. These functions have regular behavior at infinity, but they are not normalizable. Therefore, these states lie in the continuous part of the spectrum.

In the last case $\omega^2<m^2_0$, we can write
\begin{gather}
    \psi_r = 
    \begin{cases}
        aI_{k+\frac{1}{2}}\left(r\sqrt{m^2_0-\omega^2}\right)+bK_{k+\frac{1}{2}}\left(r\sqrt{m^2_0-\omega^2}\right),\, r<r_0, \\
        cI_{k+\frac{1}{2}}\left(r\sqrt{m^2_0-\omega^2}\right)+dK_{k+\frac{1}{2}}\left(r\sqrt{m^2_0-\omega^2}\right),\, r>r_0.
    \end{cases}
\end{gather}
For $k \in (-\frac{1}{2},\infty)$, the regularity condition requires us to set $b=c=0$, and then we obtain
\begin{gather}
    \psi_r = 
    \begin{cases}
        aI_{k+\frac{1}{2}}\left(r\sqrt{m^2_0-\omega^2}\right),\, r<r_0, \\
        dK_{k+\frac{1}{2}}\left(r\sqrt{m^2_0-\omega^2}\right),\, r>r_0.
    \end{cases}
\end{gather}
From the original matrix equation, we can derive the components for $\psi_{\psi}$
\begin{gather}
    \psi_{\phi} = 
    \begin{cases}
        a\sqrt{\frac{m_0+\omega}{m_0-\omega}}I_{k-\frac{1}{2}}\left(r\sqrt{m^2_0-\omega^2}\right)\, r<r_0, \\
        d\sqrt{\frac{m_0-\omega}{m_0+\omega}}K_{k-\frac{1}{2}}\left(r\sqrt{m^2_0-\omega^2}\right),\, r>r_0,
    \end{cases}
\end{gather}
where we also used the relations
\begin{gather}
    I'_{\nu}(z)=-\frac{\nu}{z}I_{\nu}(z)+I_{\nu-1}(z),\\
    K'_{\nu}(z)=-\frac{\nu}{z}K_{\nu}(z)-K_{\nu-1}(z).
\end{gather}

Therefore, using the requirement that the spinor should be continuous, we obtain the dispersion relation
\begin{equation}\label{eq:EQ}
    \frac{\omega-m_0}{\omega+m_0}\frac{K_{k-\frac{1}{2}}(X)}{K_{k+\frac{1}{2}}(X)}+\frac{I_{k-\frac{1}{2}}(X)}{I_{k+\frac{1}{2}}(X)}=0,
\end{equation}
where $X=r_0 \sqrt{m^2_0-\omega^2}$. Now, let us consider the case $r_0\gg 1$ and use the expansion for large $z$
\begin{gather}
    I_{\nu}(z)\approx \frac{e^z}{\sqrt{2\pi z}}\left(1-\frac{4\nu^2-1}{8z}\right), \\
    K_{\nu}(z)\approx \frac{e^{-z}}{\sqrt{2\pi z}}\left(1+\frac{4\nu^2-1}{8z}\right),
\end{gather}
which, after the substitution in the equation~\eqref{eq:EQ}, gives
\begin{equation}
    \frac{\omega-m_0}{\omega+m_0}\left(1-\frac{k}{r_0\sqrt{m^2_0-\omega^2}} \right)+\left(1+\frac{k}{r_0\sqrt{m^2_0-\omega^2}} \right)=0.
\end{equation}
After some algebra, the dispersion relation for the mode propagating along the circular boundary for large $r_0$ takes the form
\begin{equation}
    \omega=-\frac{k}{r_0},
\end{equation}
so it has a form similar to the flat two-dimensional boundary case.

We can also consider the smooth approximation for the change of mass sign, namely
\begin{gather}
    m=m_0 (r-r_0).
\end{gather}
In the same way, we can expand the curvature term
\begin{gather}
    \frac{k}{r}\approx \frac{k}{r_0}\left(1+\frac{r-r_0}{r_0}\right),
\end{gather}
and introduce a new variable $x=r-r_0$. In this case, functions $u_r$ and $u_\phi$ satisfy the following system of equations:
\begin{gather}\label{eq:JR_Dirac_linearappx_1}
    \partial_x \begin{pmatrix}u{_r} \\ u_{\phi}\end{pmatrix}=\begin{pmatrix}-\frac{k}{r_0}(1+\frac{x}{r_0}) & -\omega+m_0 x\\\omega+m_0 x & \frac{k}{r_0}(1+\frac{x}{r_0}) \end{pmatrix}\begin{pmatrix}u{_r} \\ u_{\phi}\end{pmatrix}
\end{gather}

it is convenient to change basis, so that the part depending on $x$ coordinate is diagonal:
\begin{gather}
    \begin{pmatrix}u{_r} \\ u_{\phi}\end{pmatrix}=\begin{pmatrix}-(\frac{k}{r^2_0}+\Omega) & -\frac{k}{r^2_0}+\Omega\\ m_0 & m_0\end{pmatrix}\begin{pmatrix}u_{-} \\ u_{+}\end{pmatrix}
\end{gather}
where $\Omega=\sqrt{\frac{k^2}{r^4_0}+m^2_0}$. In this basis, Eq. \eqref{eq:JR_Dirac_linearappx_1} takes the form:
\begin{gather}\label{eq:JR_Dirac_linearappx_2}
    \partial_y \begin{pmatrix}u_{-} \\ u_{+}\end{pmatrix}=\begin{pmatrix}-\Omega y & (\Omega-\frac{k}{r^2_0})(\frac{\omega}{m_0}+\frac{k}{r_0\Omega})\\ (\Omega+\frac{k}{r^2_0})(\frac{k}{r_0\Omega}-\frac{\omega}{m_0})& \Omega y \end{pmatrix}\begin{pmatrix}u_{-} \\ u_{+}\end{pmatrix}
\end{gather}
where we also make a shift of the coordinates $y=x+\frac{k^2}{r^3_0\Omega^2}$. From here we can follow the derivation of dispersion bands in the flat case ~\cite{Green:2020}. There is a bounded solution with dispersion relation
\be\label{eq:dispersion_sharp_linappx}
\omega=\frac{m_0 k}{r_0\sqrt{\frac{k^2}{r^4_0}+m^2_0}}
\ee
with the solution given by the first row of the Eq. \eqref{eq:JR_Dirac_linearappx_2} and condition $u_{-}=0$. If we introduce a normalized wave vector $k'=k/r_0$, in the limit $r_0 \to \infty$, we reproduce the result for the flat case and the bound mode for the sharp boundary.

There is also a series of bound states, that can propagate in both directions, which can be obtained after rewriting the Eq. \eqref{eq:JR_Dirac_linearappx_2} as a differential equation on $u_{-}$:
\be
\omega^2=\frac{k^2}{r^2_0(1+\frac{k^2}{r^4_0 m^2_0})}+(2n+2)\Omega, \;\;\; n=0,1,2,\ldots
\ee

\section{Examples of Fermi coordinates}\label{sec:fermi_cooordinates}

In many cases, it is difficult to derive the exact analytical expressions for Fermi coordinates. Nevertheless, there are cases of normal Fermi coordinates that possess an apparent formula. For example, let us consider a unit circle in a plane. Then, each straight line crossing zero is orthogonal to the unit circle, and straight lines are geodesics. Therefore, polar coordinates are Fermi coordinates for the unit circle.

Another example is a circle on a unit sphere
\begin{gather}
    \tan\theta=\cot\alpha\cos(\phi-\phi_0)+h,
\end{gather}
where $\alpha$ is a parameter corresponding to the angle between the axis of the circle and the $z$-axis, and $h$ is the distance of a circle plane from the origin of a sphere. In this case, Fermi coordinates can be constructed without difficulties. At first, we emit orthogonal geodesics, which are big circles of the sphere, and then we introduce spherical coordinates, where the two poles are the points at which these geodesics intersect. In the end, we need to normalize them so that the local frame on the curve is orthonormal and not just orthogonal. The new coordinates are related to the old ones by the rotation and rescaling
\begin{gather}
    \tan \theta'=\frac{\cos \left(\phi _0\right) (\cos (\gamma ) \cos (\theta ) \sin (\phi )-\sin (\gamma ) \sin (\theta ))+\cos (\theta ) \sin \left(\phi _0\right) \cos (\phi )}{\sin (\gamma ) \sin (\theta ) \sin \left(\phi _0\right)+\cos (\theta ) \left(\cos (\phi ) \cos \left(\phi _0\right)-\cos (\gamma ) \sin (\phi ) \sin \left(\phi _0\right)\right)}, \\
    \tan |h| \phi'=\frac{\cos \left(\phi _0\right) (\cos (\gamma ) \cos (\theta ) \sin (\phi )-\sin (\gamma ) \sin (\theta ))+\cos (\theta ) \sin \left(\phi _0\right) \cos (\phi )}{\sin (\gamma ) \sin (\theta ) \sin \left(\phi _0\right)+\cos (\theta ) \left(\cos (\phi ) \cos \left(\phi _0\right)-\cos (\gamma ) \sin (\phi ) \sin \left(\phi _0\right)\right)}.
\end{gather}


\newpage
\begin{thebibliography}{99}

\bibitem{Kane:2010}
M. Z. Hasan and C. L. Kane, ``Colloquium: Topological insulators,'' Rev. Mod. Phys. 82, (2010).

\bibitem{Moore:2010}
J. Moore, ``The birth of topological insulators,'' Nature 464, (2010).

\bibitem{Ludwig:2010}
S. Ryu, A. P. Schnyder, A. Furusaki, and A. W. W. Ludwig, ``Topological insulators and superconductors: ten-fold way and dimensional hierarchy,'' New Journal of Physics 12, (2010).

\bibitem{Irvine:2015}
Lisa M. Nash, Dustin Kleckner, Alismari Read, Vincenzo Vitelli, Ari M. Turner, and William T. M. Irvine ``Topological mechanics of gyroscopic metamaterials'', PNAS, 112, 47 (2015).

\bibitem{Xuetal:2017}
Xu, N., Xu, Y. and Zhu, J. ``Topological insulators for thermoelectrics'', npj Quant Mater 2, 51 (2017)

\bibitem{FuQin:2021}
Yichen Fu and Hong Qin ``Topological phases and bulk-edge correspondence of magnetized cold plasmas,'' Nature Communications, 12 (2021).

\bibitem{Parker:2020}
Jeffrey B. Parker, J. W. Burby, J. B. Marston, and Steven M. Tobias ``Nontrivial topology in the continuous spectrum of a magnetized plasma,'' Phys. Rev. Research 2, 033425 (2020)

\bibitem{Parker:2021}
Parker, J ``Topological phase in plasma physics.'' Journal of Plasma Physics, 87(2), 835870202 (2021)

\bibitem{Vitelli:2022}
Suraj Shankar, Anton Souslov, Mark J. Bowick, M. Cristina Marchetti and Vincenzo Vitelli ``Topological mechanics of gyroscopic metamaterials'', Nature Reviews Physics, 4 (2022).

\bibitem{Huber:2016}
S. D. Huber, ``Topological mechanics,'' Nature Physics 12, (2016).

\bibitem{PanSun:2022}
Wen-Bin Pan and Ya-Wen Sun  ``More on topological hydrodynamic modes,'' J. High Energ. Phys. 2022, 40 (2022).

\bibitem{Frey:2020}
Johannes Knebel, Philipp M. Geiger, and Erwin Frey,'' Topological Phase Transition in Coupled Rock-Paper-Scissors Cycles'', Phys. Rev. Lett. 125, 258301 (2020).

\bibitem{Delplace:2017}
Pierre Delplace, J. B. Marston and Antoine Venaille ``Topological origin of equatorial waves'', Science, 24 (2017)

\bibitem{Green:2020}
Richard Green, Jay Armas, Jan de Boer, Luca Giomi
``Topological waves in passive and active fluids on curved surfaces: a unified picture,'' arXiv:2011.12271.

\bibitem{Efimkin:2022}
G. Li and D. K. Efimkin ``Equatorial Waves in Rotating Bubble-Trapped Superfluids,'' arXiv:2210.10525.

\bibitem{Finnigan:2022}
C. Finnigan, M. Kargarian, and D. K. Efimkin ``Equatorial magnetoplasma waves,'' Phys. Rev. B 105, 205426 (2022).

\bibitem{Tong:2023}
David Tong ``A gauge theory for shallow water,'' SciPost Phys. 14, 102 (2023)

\bibitem{Nastase:2023}
Horatiu Nastase, Jacob Sonnenschein ``Euler fluid in 2+1 dimensions as a gauge theory, and an action for the Euler fluid in any dimension,'' arXiv:2303.15229.

\bibitem{Eling:2023}
Christopher Eling ``A gauge theory for the 2+1 dimensional incompressible Euler equations,'' arXiv:2305.04394 .

\bibitem{Guinea:2012}
F. Guinea,'' Strain engineering in graphene'', Solid State Communications 152 (2012).

\bibitem{Dai_etal:2019}
Zhaohe Dai, Luqi Liu, Zhong Zhang, ``Strain Engineering of 2D Materials: Issues and Opportunities at the Interface'', Adv. Mater. 31 (2019).

\bibitem{Fasolino:2007}
A. Fasolino, J. Los. \& M. Katsnelson,``Intrinsic ripples in graphene'', Nature Mater 6, 858-861 (2007).

\bibitem{nherm1} Shen, H., Zhen, B., \& Fu, L. (2018). Topological band theory for non-Hermitian Hamiltonians. Physical review letters, 120(14), 146402.

\bibitem{nherm2} Ghatak, A., \& Das, T. (2019). New topological invariants in non-Hermitian systems. Journal of Physics: Condensed Matter, 31(26), 263001.

\bibitem{nherm3} Gong, Z., Ashida, Y., Kawabata, K., Takasan, K., Higashikawa, S., \& Ueda, M. (2018). Topological phases of non-Hermitian systems. Physical Review X, 8(3), 031079.

\bibitem{Ashida:2020}
Yuto Ashida, Zongping Gong \& Masahito Ueda ``Non-Hermitian physics,'' Advances in Physics 69 (3), 249-435 (2020).

\bibitem{Shankar:2017}
S.Shankar, M.J.Bowick, and M.C.Marchetti, ``Topological sound and flocking on curved surfaces,'' Phys. Rev. X7, 031039 (2017)

\bibitem{Kunst:2019}
Flore K. Kunst, and Vatsal Dwivedi ``Non-Hermitian systems and topology: A transfer-matrix perspective,'' Phys. Rev. B99, 245116 (2019)

\bibitem{Okuma:2023}
Nobuyuki Okuma and Masatoshi Sato ``Non-Hermitian Topological Phenomena: A Review,'' 
Annual Review of Condensed Matter Physics 14(1), 83-107 (2023).

\bibitem{Bergholtz:2021}
Emil J. Bergholtz, Jan Carl Budich, and Flore K. Kunst, ``Exceptional topology of non-Hermitian systems'' Rev. Mod. Phys. 93, 015005 (2021).

\bibitem{Manasse:1963}
F.K. Manasse and C.W. Misner, ''Fermi normal coordinates and some basic concepts in differential geometry,'' Journal of mathematical physics, 4(6), 735-745 (1963).

\bibitem{Dey:2014}
Anshuman Dey, Abhisek Samanta, and Tapobrata Sarkar, ''Fermi normal coordinates and fermion curvature couplings in general relativity,'' Phys. Rev. D 89, 104008 (2014)

\bibitem{Delplace:2022}
L. Jezequel, P. Delplace ``Non-Hermitian spectral flow and Berry-Chern monopoles'', arXiv:2209.03876.

\bibitem{Rossi:2023}
Sylvain Rossi, Alessandro Tarantola ``Topology of 2D Dirac operators with variable mass and an application to shallow-water waves'', arXiv:2307.07548.

\bibitem{Tauber:2019}
C. Tauber, P. Delplace, and A. Venaille ``A bulk-interface correspondence for equatorial waves,'' Journal of Fluid Mechanics 868 (2019).

\bibitem{Murakami:2016}
K. Yokomizo and S. Murakami, ``Non-Bloch Band Theory of Non-Hermitian Systems'' Phys. Rev. Lett. 123, (2019).

\bibitem{Kawabata:2019}
K. Kawabata, K. Shiozaki, M. Ueda, and M. Sato, ``Symmetry and Topology in Non-Hermitian Physics'' Phys. Rev. X 9 (2019).

\bibitem{Yao:2018}
S. Yao and Z. Wang, ``Edge States and Topological Invariants of Non-Hermitian Systems'' Phys. Rev. Lett. 121 (2018).

\bibitem{Schindler:2023}
Frank Schindler, Kaiyuan Gu, Biao Lian, and Kohei Kawabata, ``Hermitian Bulk -- Non-Hermitian Boundary Correspondence'' arXiv:2304.03742.

\bibitem{Delplace:2020}
M. Marciani and P. Delplace ``Chiral Maxwell waves in continuous media from Berry monopoles'', Phys.Rev. A 101, 023827 (2020).

\bibitem{Mechelen:2018}
Todd Van Mechelen and Zubin Jacob ``Quantum gyroelectric effect: Photon spin-1 quantization in continuum topological bosonic phases,'' Phys. Rev. A 98, 023842 (2018) in Differential Geometry," J. Math. Phys. 4, 735 (1963).

\bibitem{Bal:2019}
G. Bal, ``Continuous bulk and interface description of topological insulators,'' Journal of Mathematical Physics 60, 081506 (2019).

\bibitem{HuXieZhu:2022}
Pipi Hu, Peng Xie, Yi Zhu ``Traveling edge states in massive Dirac equations along slowly varying edges'', arXiv:2202.13653.

\bibitem{Drouot:2022}
Alexis Drouot ``Topological insulators in semiclassical regime'', arXiv:2206.08238.

\bibitem{Gallerati:2019}
Gallerati, A. ``Graphene properties from curved space Dirac equation.'' Eur. Phys. J. Plus 134, 202 (2019).


\end{thebibliography}
\end{document}